\documentclass[a4paper,fleqn,usenatbib]{mnras}

\usepackage[T1]{fontenc}
\usepackage{ae,aecompl}

\usepackage{graphicx}	
\usepackage{amsmath}	
\usepackage{amssymb}	
\usepackage{ulem}

\def \psr {PSR J0737-3039A/B}
\def \psrb {PSR J0737-3039B}

\def \bfig {\begin{figure}}
\def \efig {\end{figure}}
\def \fig  {\includegraphics}
\def \btab {\begin{table}}
\def \etab {\end{table}}
\def \btabb {\begin{tabular}}
\def \etabb {\end{tabular}}

\def \phim2 {$\phi_{m2}$}
\def \T0   {$T_0$}

\def \los {{\it los}}
\def \nct {\nocite}

\def \mref#1{(\ref{#1})}

\title[Fan beam model for  PSR J0737 -- 3039B]
{The fan beam model for the pulse evolution of PSR J0737-3039B}

\author[L.~ Saha and J.~Dyks]{
L.~Saha\thanks{E-mail: labsaha@ncac.torun.pl}
and
J.~Dyks\\
Nicoulas Copernicus Astronomical Center, Polish Academy of Sciences, Rabia\'{n}ska 8,  87-100, Torun, Poland\\
}

\date{Accepted XXX. Received YYY; in original form ZZZ}

\pubyear{2016}

\begin{document}
\label{firstpage}
\pagerange{\pageref{firstpage}--\pageref{lastpage}}
\maketitle

\begin{abstract}
Average radio pulse profile of a pulsar B 
in a double pulsar system \psr\ 
exhibits an interesting behaviour.
During the observation period between 
2003 and 2009, the profile evolves from a single-peaked to 
a double-peaked form, following disappearance in 2008 indicating 
that the geodetic precession of the pulsar is a possible origin of 
such behaviour. The known pulsar beam
models can be used to determine 
the geometry of \psrb\ in the context of the precession.
We study how the fan-beam geometry
performs in explaining the
observed variations of the radio profile morphology. It is shown that 
the fan beam can successfully reproduce the observed evolution 
of the pulse width, and should be considered as a serious alternative for
the conal-like models. 

\end{abstract}



\section{Introduction}
\psr\ is a system of two neutron stars in a highly 
relativistic orbit discovered in 2003  by the 64 m Parkes radio telescope 
\citep{Burgay_2003, Lyne_2004}. It is an ideal example of eclipsing 
double pulsar system which provides a unique opportunity to test theories 
of general relativity \citep{Kramer_2006}. The system consists of two pulsars, 
one with a short spin period of 23 ms (hereafter pulsar ``A") and the second 
one with a longer spin period of 2.8 s (hereafter pulsar ``B"). These two 
pulsars orbit each other in a 2.4 hr cycle. The orbital inclination was 
calculated to be 88.$^\circ$7 \citep{Kramer_2006, Ransom_2004} and 
89.$^\circ$7 \citep{Coles_2005}  through two different methods establishing 
that the orbital plane is almost edge-on to our line of sight (hereafter \los), 
which is an ideal geometry to form an eclipsing binary system. Radio 
observation of this system displays a roughly 30 s eclipse of pulsar A 
by pulsar B when the emission from pulsar A is  obscured 
by the fast-spinning inner magnetosphere of pulsar B \citep{Lyne_2004,Lyutikov_2005}. 
The two pulsars are separated by a distance of about $9 \times 10^5$ km 
in their orbits as measured by radio timing analysis \citep{Lyne_2004}. 

From the prescription given in \cite{Baker_1975}, the precession rate is obtained as $5\degr.0734  \pm  0\degr.0007$/yr \citep{Breton_2008}. 
Following  the observational results of the $\sim$ 30 s eclipse of A by B, a simple geometric model was used to explain the observed morphology and to measure the geometry of B \citep{Breton_2008}. Based on the observed radio profile from pulsar A, the relativistic precession of pulsar B's spin axis around the total angular momentum was estimated to be 4\degr.77$^{+0^\circ.66}_{-0^\circ.65}$/yr \citep{Breton_2008}.

It has been found that the observed pulse profile of A is quite stable. 
However, pulse shape and the intensity of the emission from pulsar B vary 
with orbital longitude. Radio observation by Parkes telescope system during 
observation made between 2003--2005, showed two bright phases centered around 
longitude of $\sim~210^\circ$ (hereafter bp1) and $\sim~280^\circ$ (hereafter 
bp2). In addition, in bp2 the pulse profile of B was observed to evolve from a single-peaked to a double form 
\citep{Burgay_2005}.  
The presence of two bright phases 
is also established by radio observations with the 100 m Green Bank Telescope 
(GBT) made during 2003--2009 (\citealt{Perera_2010}, hereafter PMK10).
Although there was some 
overlap between the observations made by these two telescopes, the behaviour 
of the observed pulse profile is quite different. In this overlapping time 
interval, pulse profiles from the GBT system were single-peaked in both the 
bright phases. However, the profiles 
then got broader and split into a double form,
along with a reduction of their flux densities.  Finally, the pulses disappeared 
in March 2008. 

To explain the observed pulse profile and to determine the geometry 
of precessing pulsars, the 
traditional choice of emission model 
is a circular hollow-cone beam \citep{Kramer_1998} as  
initially used for PSR B1913+16. 
The profile of this pulsar, when mapped onto the plane of spin and 
precession phase, creates a pattern that can be interpreted
with an hourglass-shaped beam \citep{Weisberg_2002}. However, 
\citet{Clifton_2008} 
have shown that for a specific choice of  model parameters
the conal beam can also produce such an hourglass-shaped pattern.
Similarly, in order to explain the observed secular variation of the pulse 
profile from the \psrb\ and its disappearance in 2008, simple circular hollow-cone 
beam and elliptical horseshoe beam models are used (PMK10).
\nct{Perera_2010}
 \citet{Lyutikov_2014} have studied the secular and orbital
visibility of pulsar B, by considering the effect
of the magnetized wind from pulsar A on the magnetosphere of pulsar B. 
The basic framework of all of these studies was  the conal beam model.        

In this paper, we attempt to understand the observed changes in the pulse 
profile of pulsar B within the framework of the fan beam model
\citep{Michel_1987,Dyks_2010,Wang_2014},
which is successful in explaining various 
observational data, e.g.~double notches in pulsar profiles, 
bifurcated emission components 
\citep{Dyks_2012}, the rate of the radius to frequency mapping
\citep{Karastergiou_2007,Chen_2014, Dyks_R_2015},  
and localised  
distortions of polarisation angle curve \citep{Dyks_2016}. 

With regard to the precession-driven change of profiles, 
a slow quasi-linear evolution of the peak-to-peak 
separation had been shown to be typical of the fan beam model
 (Fig.~17 in sect.~6.3.3 of \citealt{Dyks_2010}), 
as opposed to the conal model, in which the passage through different
regions of a cone usually implies nonlinear changes.
Based on this a fan beam was predicted for the main pulse 
of PSR J1906$+$0746, and confirmed to be like that two years later 
\citep{Desvignes_2012}. 
 The interpulse emission of PSR J1906$+$0746, however, was mapped into
a non-elongated patch.

The paper is organized as follows: 
 In Section \ref{sec:model}, we set up the formulae to calculate the profile 
width \textit{W} 
based on the fan-beam model.
Section \ref{data} describes the peak-separation data used in the modelling.
 Results are presented in Section \ref{results}
 and discussed in Section \ref{sec:discussion}.

\section{Model description}\label{sec:model}
We consider a coordinate system as described in \cite{Kramer_1998} 
(see Fig.~4  therein). The angle between 
pulsar spin axis and  the line of sight, 
$\zeta(t)$, can be written as

\bfig
\centering
\fig[scale = 0.5]{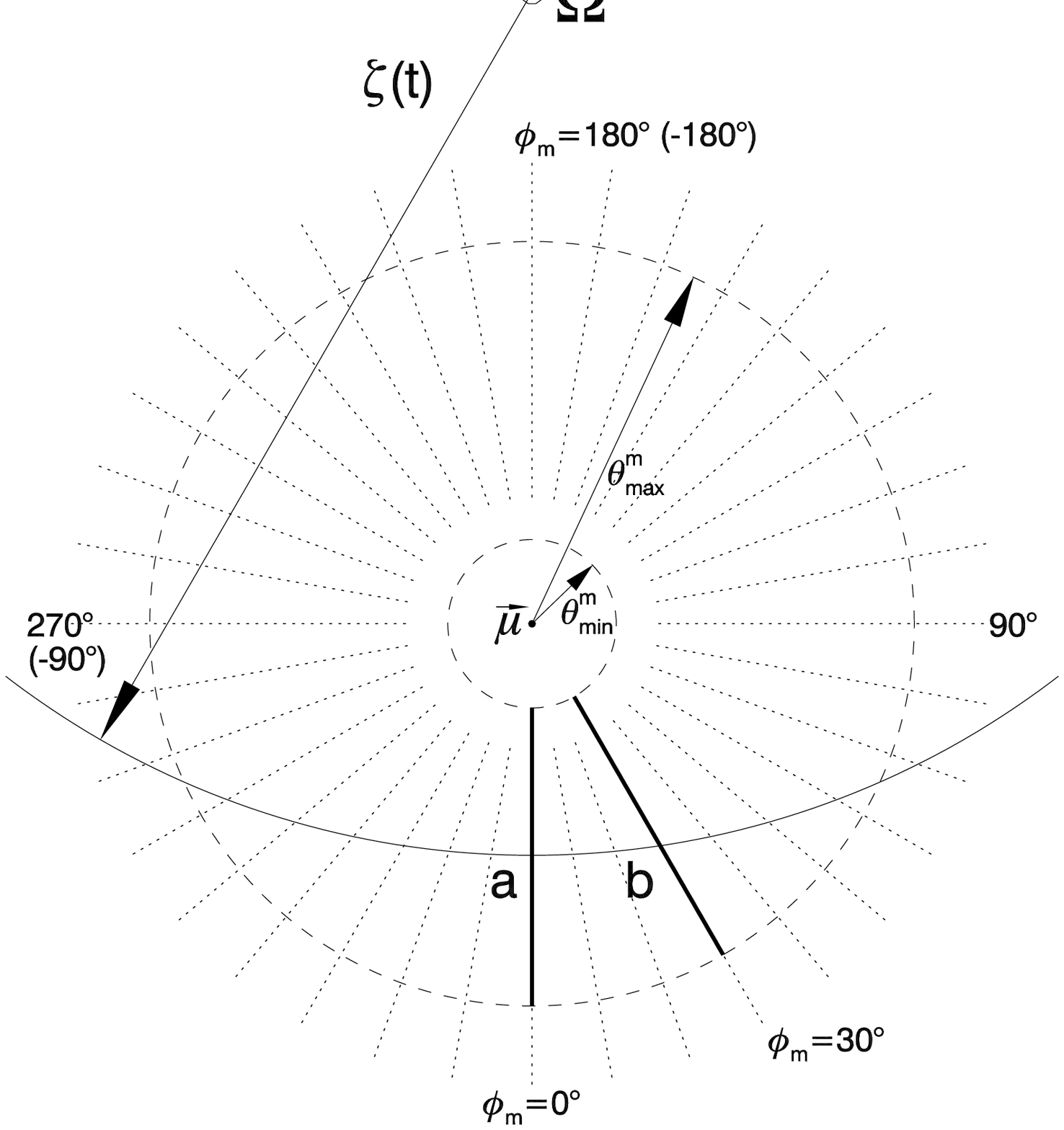}
\caption{Sky projection of dipolar magnetic field lines (dotted)
viewed from above the polar region. Two thick line segments that 
follow a fixed magnetic azimuth represent fan beams emitted by plasma streams. 
The line of sight is traversing the beams at two different 
places (e.g., a and b). The angular distance between these two 
points, as measured around $\vec \Omega$, is the observed pulse width $W$.}
\label{fig:model}
\efig

\begin{figure*}
\centering
\fig[scale = 0.88]{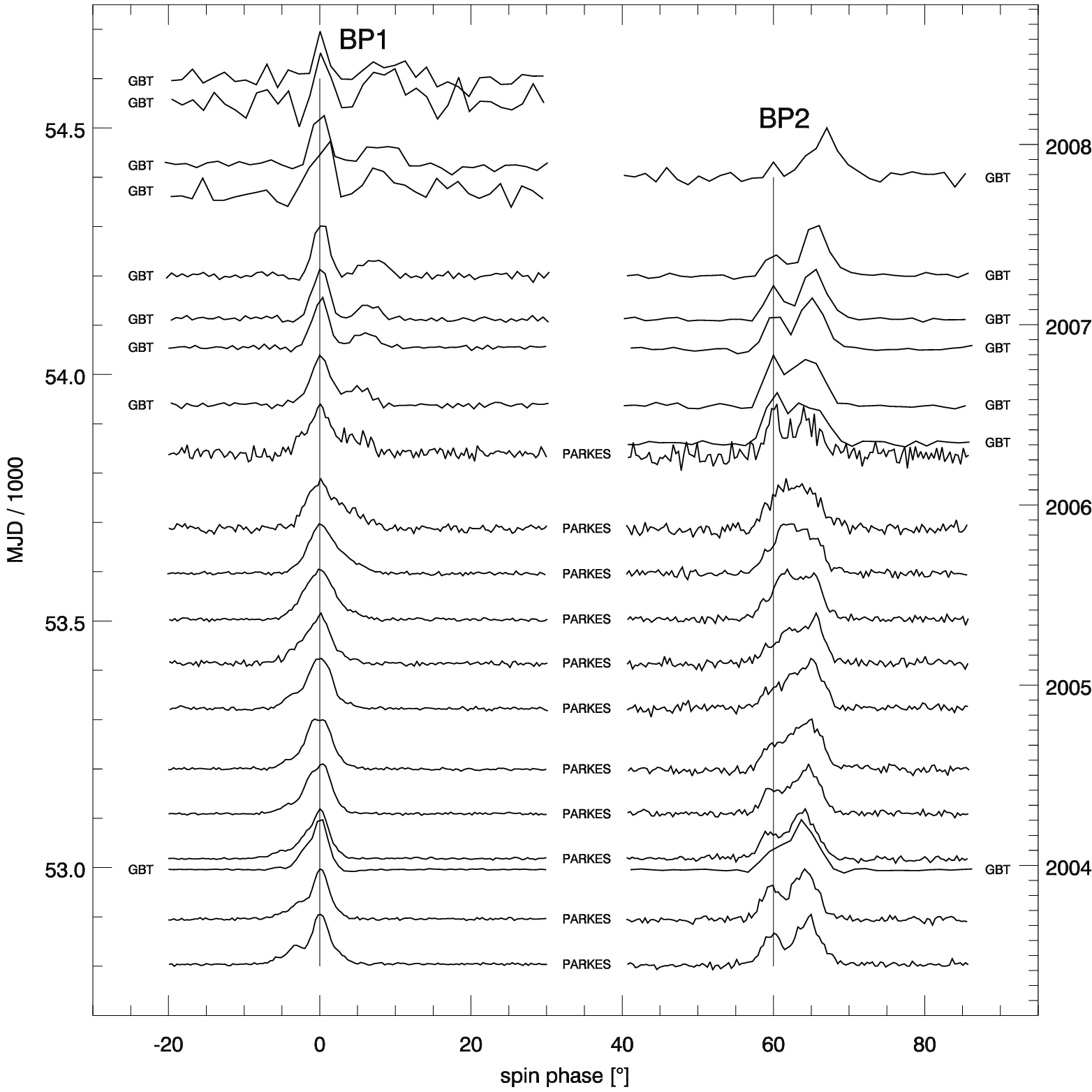}
\caption{Evolution of PSR J0737-3039B profile shown for 
both radio-bright orbital phases: bp1 (left) and bp2 (right). 
It is a compilation of published data from GBT (820 MHz, Perera et al.~2010)
and the Parkes telescope (1390 MHz, Burgay et al.~2005). 
Late time profiles 
(${\rm MJD} > 53800$ are aligned according to the leading component. 
The early time data are aligned differently for the two orbital phases:
in the case of bp1, according to the brightest peak in the profile;
in the case of bp2, to match the overall on-pulse window. 
Vertical lines are added for reference. The zero point of spin phase axis 
is arbitrary. 
Data courtesy: B.~Perera, M.~McLaughlin (GBT) 
and M.~Burgay, R.~Manchester (Parkes).
}
\label{figdata}
\end{figure*}

\begin{eqnarray}
\cos \zeta (t) = \cos \lambda \cos i + \sin \lambda \sin i \cos \Phi(t),
\label{zeta}
\end{eqnarray}
where $\lambda$ is the angle between orbital momentum and spin axis, 
$i$ is the angle of inclination of the orbital plane  normal
with respect to  the 
line of sight and $\Phi(t)$ is the precession phase which is defined as

\begin{eqnarray}
\Phi(t) = \Omega_p~(T_0-t),
\end{eqnarray}
 where $\Omega_p$ is the precession rate and  $T_0$ is the initial reference time defined as the time when spin axis 
of the pulsar comes closest to the line of sight.  
 
It is assumed that the 
radio emission pattern of J0737$-$3039B consists 
of two fan-shaped beams, each one following 
a fixed magnetic azimuth $\phi_{m,i}$ with $i\in\{1,2\}$
(thick-line sections in Fig.~\ref{fig:model}). 
To minimise the number of free parameters, two
 cases are considered in which one of the beams (with the index `1') 
is located in a specific way. In an asymmetric case,  the beam 1 
is oriented meridionally 
(either at $\phi_{m,1}=0$ or $\phi_{m,1}=\pi$). In a symmetric case, 
both beams are equidistant from the main meridian
($\phi_{m,1}=-\phi_{m,2}$).
The line of sight is cutting through the fan beams at a magnetic colatitude
$\theta_m$, which can be calculated from the following equation:
\begin{eqnarray}
\cos \zeta (t) = \cos (\pi - \phi_m) \sin \alpha \sin \theta_m 
+ \cos \alpha \cos \theta_m, 
\label{eqn:zeta2}
\end{eqnarray}
where $\alpha$ is the angle between the magnetic axis $\vec \mu$
and the pulsar spin axis $\vec \Omega$ 
(for details see the Appendix A in \citealt{Dyks_2015}).

The pulse longitude, $\phi$, for that point where the magnetic azimuth $\phi_m$ 
and 
$\zeta(t)$ intersect each other at time $t$ can then be written as
\begin{eqnarray}
\cos \phi(t) = {\cos \theta_m - \cos \alpha \cos \zeta (t) 
\over \sin \alpha \sin \zeta (t)}.
\label{eqn:phi}
\end{eqnarray}

The pulse width $W$ is defined here as the separation between the peaks of 
observed components, and the peaks are associated with 
the locations of the fan beams in Fig.~\ref{fig:model}. 
In the conal-beam model, 
$W$ can be written as $2\phi(\theta_m, \alpha, \zeta)$, 
where $\theta_m$ is considered as  a half of
a fixed opening angle of the conal-beam. 
In the fan beam model, however, \textit{W}  
is determined by the beams' magnetic azimuths: 
$W(t) = |\phi_2(\phi_{m2}) - 
\phi_1(\phi_{m1})|$, where 
the pulse longitudes 
$\phi_1$ and $\phi_2$ 
are calculated from 
equation \ref{eqn:phi}.

For different choices of $\alpha$, 
$\lambda$, $T_0$, 
and $\phi_{m2}$, we  
calculate
the width as a 
function of time 
and compare it to
 the observed pulse profile of pulsar B.
The range of $\alpha$ and $\lambda$ is $(0,\pi)$, whereas $T_0$ covers
full precession period.

Narrow two-peaked profiles can only be observed when both  streams 
extend
either towards the rotational equator, or away from it
 (i.e., down or up in Fig.~\ref{fig:model}).
Therefore, to probe the full range of $\phi_{m2}$ we consider
two different cases: the outer-traverse case ($\phi_{m1}=0$, 
$0<\phi_{m2}<90^\circ$) with the sightline detecting the beams at 
$\zeta > \alpha$, and the inner-traverse case ($\zeta < \alpha$, \ 
$\phi_{m1}=\pi$, \ $90^\circ<\phi_{m2}<180^\circ$, see Fig.~\ref{fig:model} 
for definition of the magnetic azimuth $\phi_{m}$). The parameters are  considered to have grid size of $0.^\circ6$ for $\alpha$ and $\lambda$,  $1^\circ$ for $\phi_{m2}$, and 1 yr for $T_0$.

When the pulsar is viewed in the equatorward region 
($\sin\zeta(t)>\sin\alpha$)
each magnetic azimuth is cut by the
sightline one time per rotation period $P$, and 
no constraints on the range of radio-bright colatitudes are imposed, 
i.e.,~$\theta^m_{\rm min}=0$ and $\theta^m_{\rm max}=\pi$.
For circumpolar viewing, however, the sightline can cut through each B-field
line twice, so two streams can produce four-peaked pulse profile. 
In principle such models may be rejected as having too many peaks,
however, since one solution for $\theta_m$ is usually much
larger than the other, and the radio emissivity is likely decreasing with
$\theta_m$, it is also possible to ignore the far value.
We choose this last option, i.e.,~whenever the beam's magnetic azimuth 
is cut
 twice per $P$, only the smaller value of $\theta_m$ is used.

Since the pulsar B has not been detected after March 2008, we reject all
models which predict detectable flux in the years 2009--2016. This is done
 in the following way. For each model with 
a given $\lambda$ and $T_0$, 
we first determine the range of viewing angle $\zeta$ which corresponds 
to the period of B's visibility (2003--2008). Then we check if any
$\zeta$ from this range occurs in the years 2009-2016. If it does, the model
is rejected. In all the following calculations,
 we use $i = 88^\circ.7$ and $\Omega_p = 4.^\circ8$/yr.

\section{Observational data}
\label{data}

Observations of PSR J0737$-$3039B have been 
described both in \citet{Burgay_2005} and PMK10,
\nct{Perera_2010} 
and we reproduce them in Fig.~\ref{figdata}.
Unfortunately,
only the late time data on peak separation from PMK10 
can be interpreted uniquely,  
and consistently for both bp1 and bp2.  
Therefore, as the main modelling target we use the GBT data recorded
between
 December 24, 2003, and June 20, 2009 (PMK10).
\nct{Perera_2010}
 As can be seen in Fig.~\ref{figdata}, the pulse 
profile was single-peaked in late 2005.
 In the early 2006 the two-peaked 
pulse profile appears and the pulse width increases over time 
 while the flux is decreasing. 
Finally, the radio signal of pulsar B 
disappears at the end of March 2008. 
This characteristic has been seen for both the phases, bp1, and bp2.

We use the data on 
 the pulse width, i.e.,~the peak-to-peak separation of components, 
as determined in PMK10, 
although we do not use the same errors.
The changes of $W$, illustrated in Fig.~6 of PMK10, 
can be well described as linear except
their dispersion around the linear trend is much larger than
expected for the Gaussian distribution. A likely reason for this is the 
low time resolution of the data (512 or 256 bins across the spin period).
There are sometimes only four data points within each profile component, 
which is fitted with a three-parameter Gaussian at only one degree of freedom. 
The number of flux measurements which are involved in the fitting, is then 
much smaller than the number of those used to
calculate their one-sigma error (several tens of data points within 
the noisy off-pulse region).  
When the errors of $W$ are 
rescaled by a factor of $1.7$, then 68 per cent of data points 
is within the reach of the linear trend. 
Therefore, in the following, we assume that 68\% confidence errors 
(hereafter called $1\sigma$) 
are larger by a factor of $1.7$ than those published in PMK10.

\section{Results}
\label{results}

\bfig
\centering
\fig[scale = 0.40]{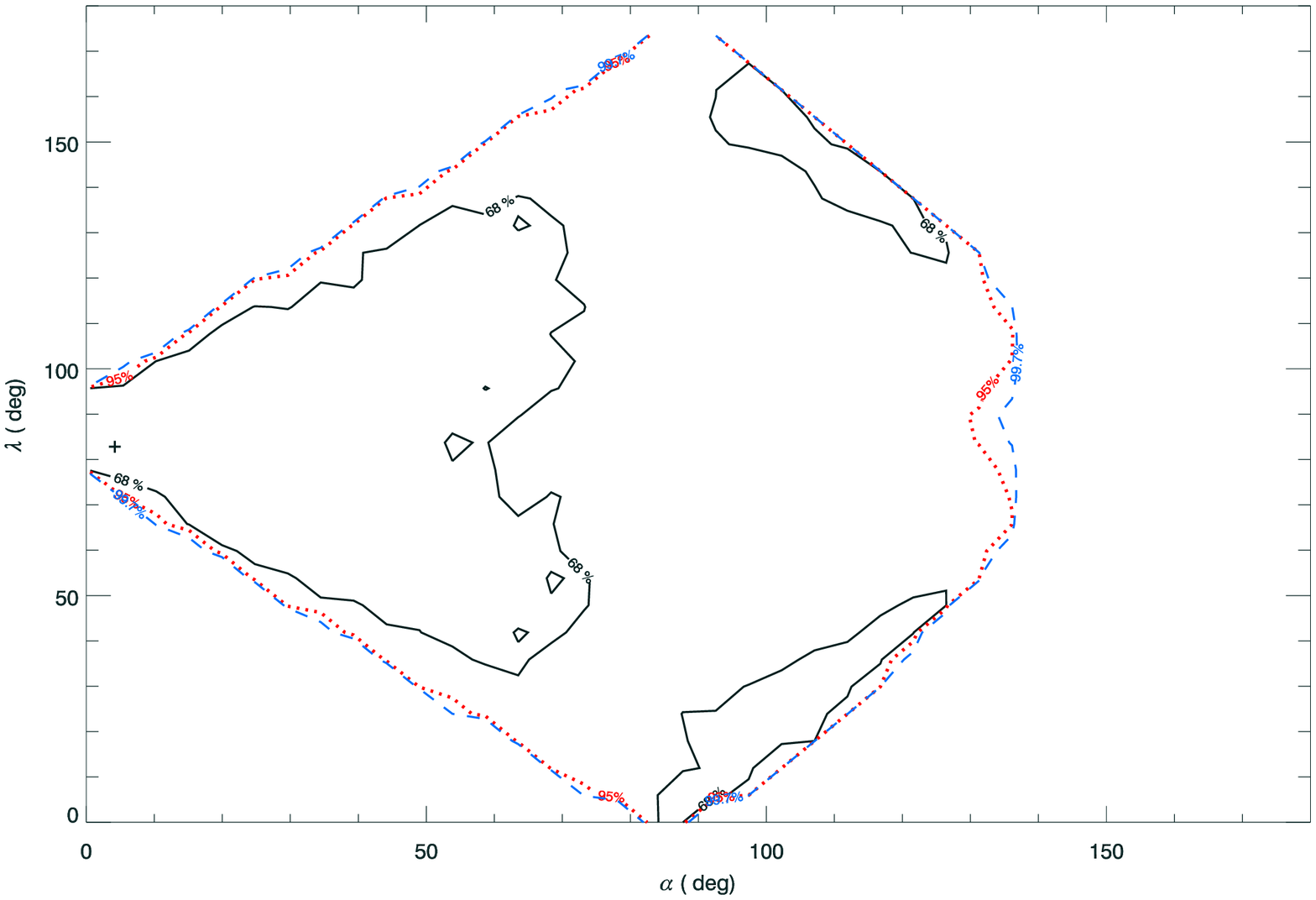}
\caption{$\chi^2$ map for the fan beam model of the pulse width data 
on pulsar B. 
The 
$\chi^2$ contours are for the confidence levels of 1$\sigma$ (solid line), 
2$\sigma$ (dotted line) and 3$\sigma$ (dashed line). 
The `plus' sign on the left marks the minimum $\chi^2$ value, 
 which should not be considered as a meaningful best-fit solution. 
The result was obtained for the outer traverse through the asymmetric
beam ($\phi_{m1}=0$).
}
\label{fig:contour_case3}
\efig

Fig.~\ref{fig:contour_case3} presents the 1$\sigma$, 2$\sigma$ and 
3$\sigma$ contours of $\chi^2$ on the $(\alpha, \lambda)$ plane, as
calculated for the asymmetric outer case ($\phi_{m1}=0$, $\phi_{m2} <
90^\circ$). 
It can be seen that for a very large range of $\alpha$ and $\lambda$ 
the model can reproduce the data with accuracy better than $2\sigma$.
The pronounced bay of high $\chi^2$ on the right-hand side is caused 
by the condition of invisibility of pulsar B in the years 2009-2016,  
which resulted in the rejection of some good-precision fits. Without the
invisibility condition, the $2\sigma$ and $3\sigma$ contours of $\chi^2$ 
have the shape of a undistorted rhombus.

The observations constrain both the observed width $W_{\rm obs}(t)$ 
and the roughly constant rate of
width change $(dW/dt)_{\rm obs}$ at 
any moment when the pulsar B is detectable. 
The high capability of the model to adjust to both $W_{\rm obs}$ 
and $(dW/dt)_{\rm obs}$
can be most easily 
 understood in the case of an orthogonal precession ($\lambda =
90^\circ$, middle horizontal region of Fig.~\ref{fig:contour_case3}).

Since the observed profile
 is only a few degrees wide, 
our sightline must traverse the beam close to the dipole axis.
 Stated otherwise, the sightline's 
smallest  angular distance from the dipole axis, 
as measured within a given pulsar spin period, i.e.,~the impact angle 
$\beta=\zeta-\alpha$, must be generally small
($\beta \ll \alpha \sim \zeta$).\footnote{ Special cases, like those 
with $\phi_{m1}\approx\phi_{m2}$ can give small $W$ even at large $\beta$, 
however, these cases constitute a marginal periphery of 
the whole parameter space, and usually do not match $dW/dt$. 
Typically, within
that part of the parameter space which can possibly reproduce the data,
 the impact angle $\beta$ needs to be small.}
The path of a sightline traverse through the beam may then be considered
a straight line orthogonal to the main meridian, and 
the pulse width $W$ can be roughly approximated by:
\begin{equation}
W\approx\beta\tan\phi_{m2}/\sin\zeta\approx(\zeta(t)
-\alpha)\tan\phi_{m2}/\sin\alpha.
\label{wapp}
\end{equation}
Note that the small difference between $\zeta$ and $\alpha$ has been neglected
in the denominator which takes into account the `small circle' effect.
Accordingly, the
derivative of $W$ is
\begin{equation}
\frac{dW}{dt} \approx \frac{-\Omega_p \tan\phi_{m2} \sin\lambda \sin i 
\sin \Phi(t)}{
\sin\alpha\sin\zeta(t)}.
\label{deriv}
\end{equation}
In the case of the orthogonal precession ($\lambda\approx 90^\circ$), 
the near-orthogonal orbit inclination $i$ 
implies,
through eq.~\mref{zeta}, that  $|\Phi(t)| \approx |\zeta(t)|$
and eq.~\mref{deriv} reduces to $|dW/dt| \approx |\Omega_p \tan\phi_{m2}|/\sin\alpha$. 
Thus for any $\alpha$ the model can reproduce an arbitrary value of $|dW/dt|$ 
through adjustment of the azimuthal beam separation 
($\phi_{m2}$). In this orthogonal case ($\lambda\sim i\sim90^\circ$),
 the width given by eq.~\mref{wapp} 
is simply equal to $W\approx |[\Omega_p (T_0-t)-\alpha]
\tan\phi_{m2}|/\sin\alpha$
and for any set of $(\alpha, \phi_{m2}, t)$ the width can be 
reproduced by adjustment of the impact angle through $T_0$. 

\bfig
\centering
\fig[scale = 0.40]{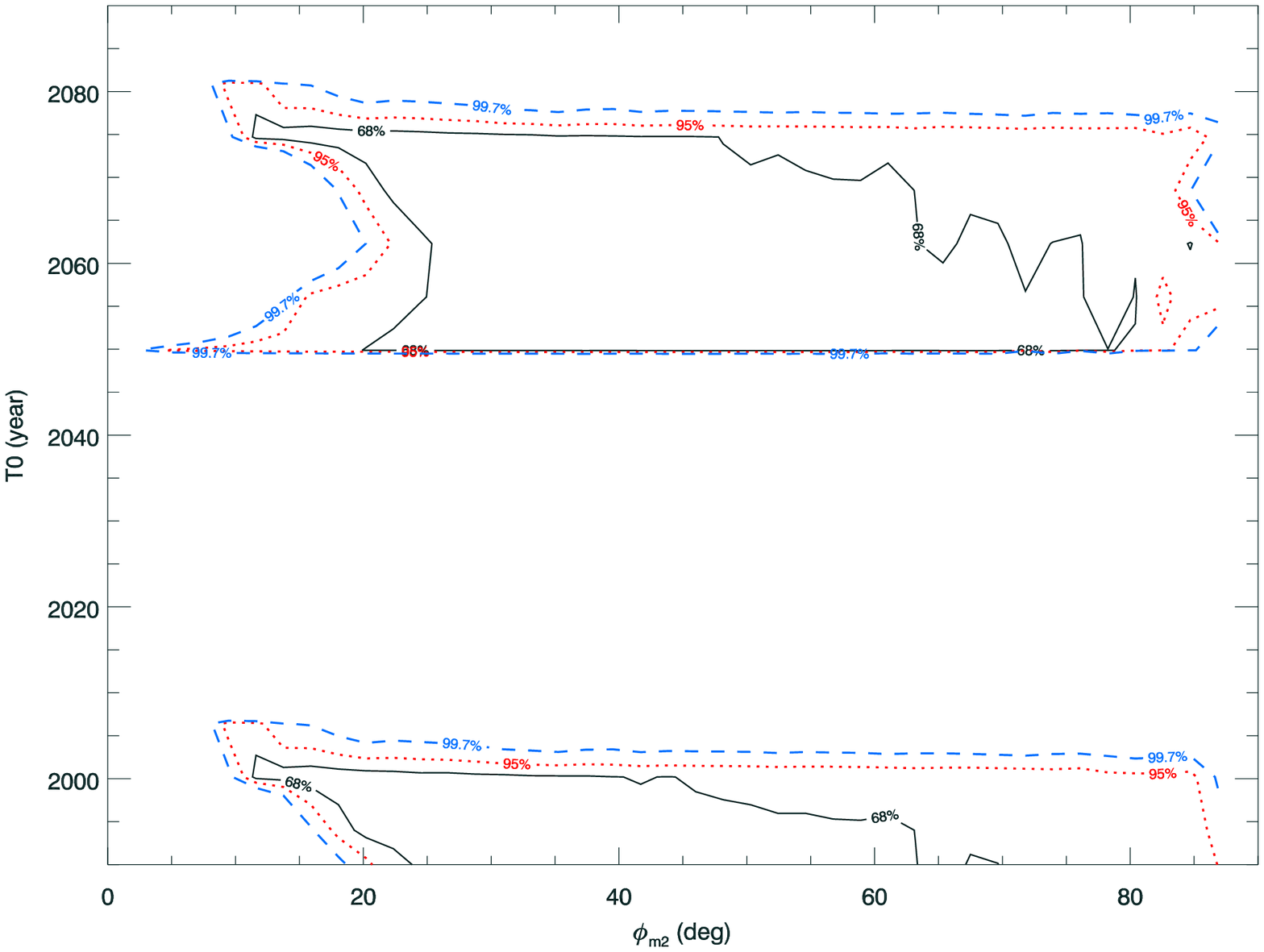}
\caption{The map of $\chi^2$ on the $(\phi_{m2}, T_0)$ plane, calculated
for the same case, and the same confidence levels as in the previous
figure.
}
\label{phim2t0}
\efig

In the corners of Fig.~\ref{fig:contour_case3}, no combination of model
parameters can reproduce the data. This can be understood by considering the
case of 
$\lambda\ll 1$ rad
and $\alpha \ll 1$ rad. The dipolar magnetosphere is then viewed roughly 
at a right angle with respect to the dipole axis 
($\zeta\sim\theta_m \sim 90^\circ$) which implies 
$W \approx \phi_{m2}$ and $dW/dt \approx 0$.
Thus, for the small precession angle the precession cannot ensure 
as steep variation of $W$ as observed. At a fixed $\alpha$, the value 
of $\lambda$ needs to be sufficiently large for the observed $dW/dt$
to appear at some precession phase. The parameters ($T_0$, $\phi_{m2}$) 
are then adjusted to allow  the model to match the data. Beyond the rhombus region this is
not possible.

The small-angle precession ($\lambda \ll 1$ rad) can only reproduce
the data when $\alpha\approx 90^\circ$ (bottom and top corners of the
rhombus). This is because 
the beam may now be viewed arbitrarily close to the dipole axis, so 
the tiny wiggling of the magnetosphere can change the small value of 
$\beta$ by a large factor.

It needs to be emphasized that the symmetric rhombus shape of the
low-$\chi^2$ contours is characteristic of the orthogonal orbit inclination
with respect to the line of sight ($i\approx 90^\circ$). For $i\ne90^\circ$,
the rhombus gets transformed into an elongated parallelogram.

The confidence contours ($68$, $95$ and $99.7$\%) are also shown on the 
$(\phi_{m2}, T_0)$ map (Fig.~\ref{phim2t0}). Again the observed $W$ can 
be reproduced with reasonable precision within a very large 
part of parameter space. There are horizontal bands visible, in which
 $T_0$ gives no acceptable solution. In these bands $\zeta$ changes with time 
in a wrong direction, causing $W$ to decrease with time, in contrast to the 
observations. The bands have the width of a half precession period
and their borders do not depend on $\phi_{m2}$ because the
 evolution of $\zeta$ is fully determined by $\lambda$, $i$ and $T_0$ 
(eq.~\ref{zeta}).

\bfig
\centering
\fig[scale = 0.4]{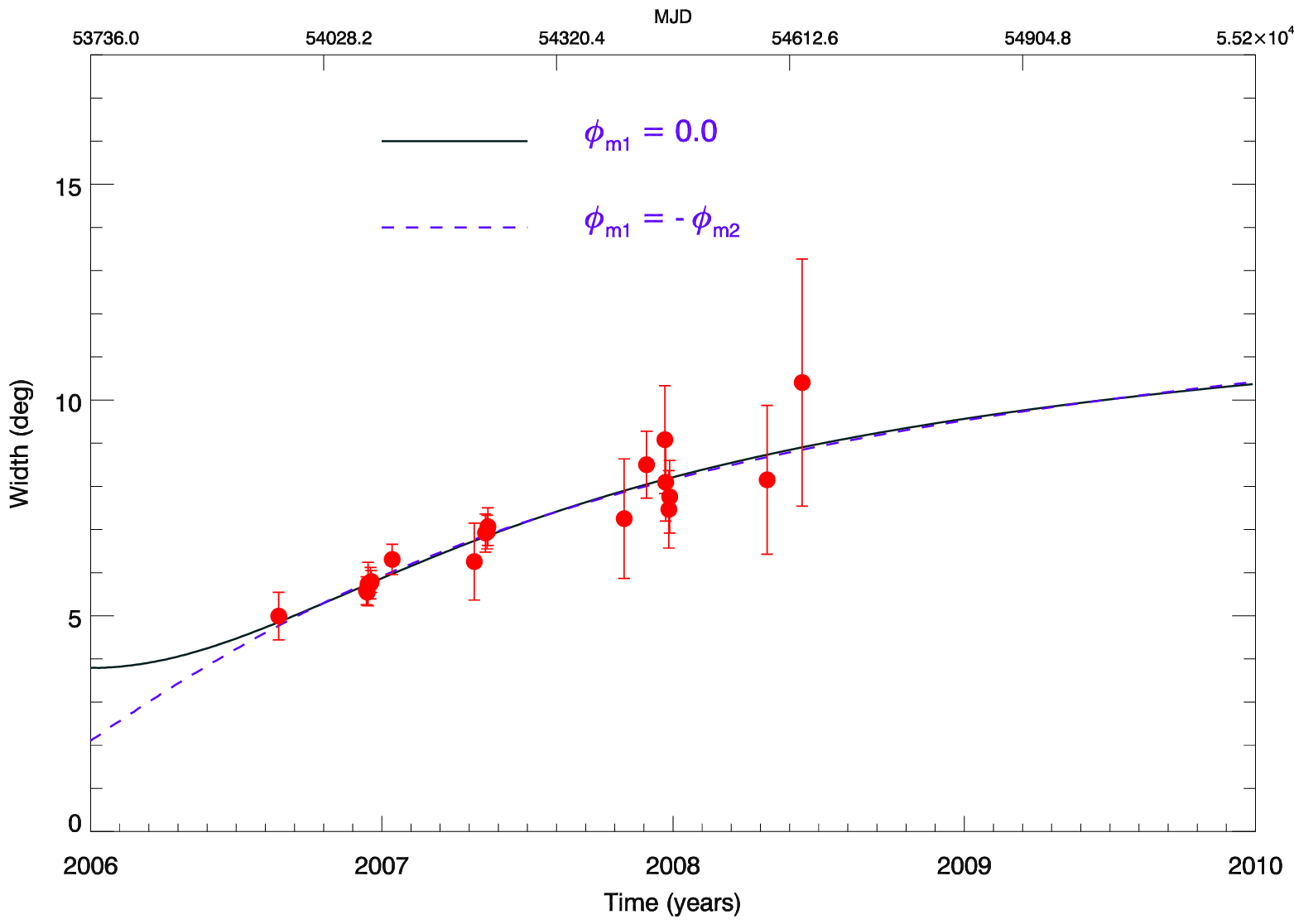}
\caption{Nominally-best fit of the fan beam model to the pulse width
 data 
taken with GBT between  Dec 24, 2003 
and Jun 20, 2009 (PMK10).  
Solid line presents the asymmetric beam case marked with the cross 
in Fig.~\ref{fig:contour_case3} 
($\phi_{m1} = 0$). 
}
\label{fig:model-fit}
\efig

\bfig
\centering
\fig[scale = 0.40]{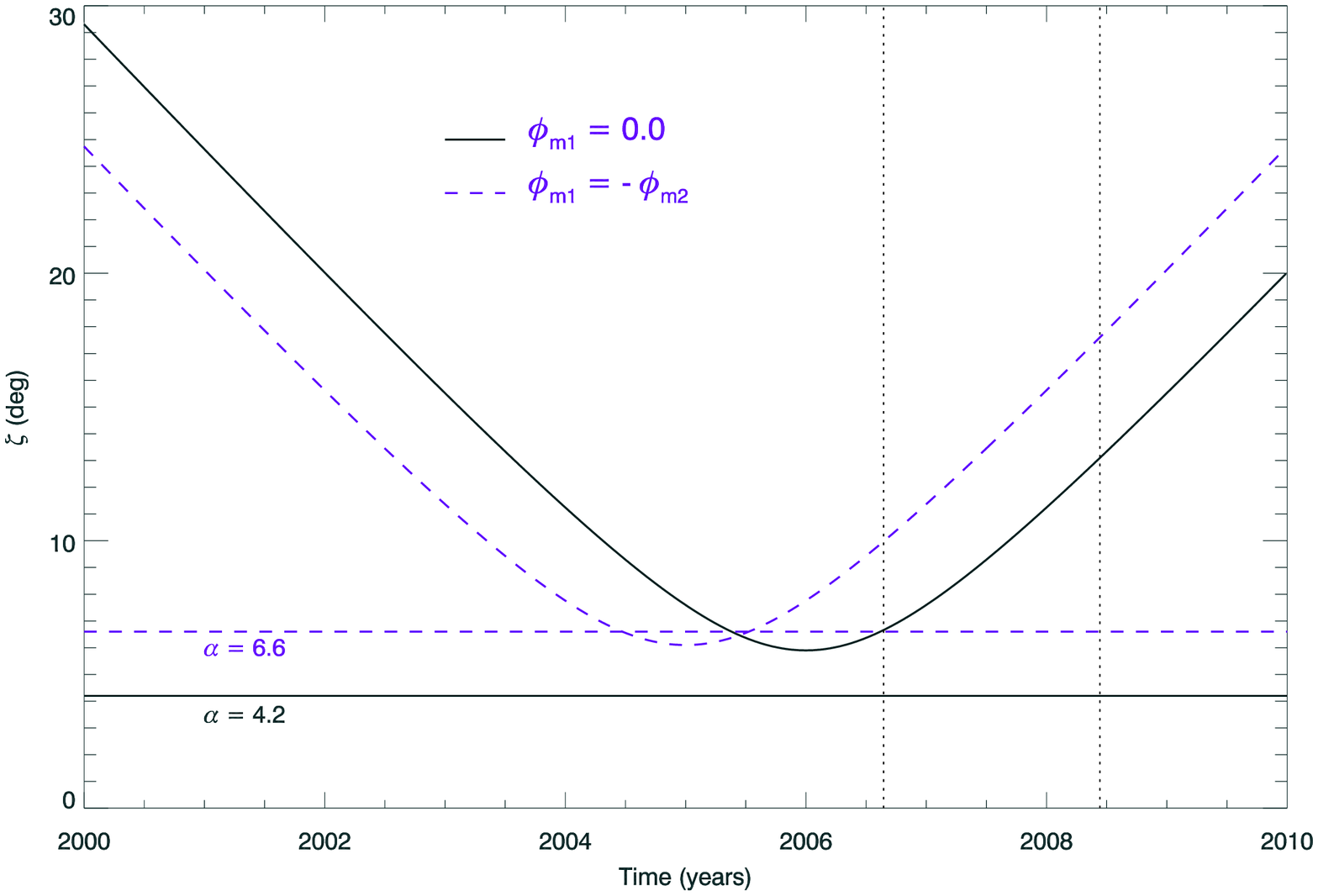}
\caption{ The viewing angle $\zeta$ as a function of time for
the models shown in the previous figure (solid line: $\phi_{m1}$ = 0, 
dashed line: $\phi_{m2} = - \phi_{m1}$). 
Two vertical lines are drawn to indicate the region where the
peak-separation data is available 
from the GBT.}
\label{fig:zeta_case1}
\efig

\bfig
\centering
\fig[scale = 0.40]{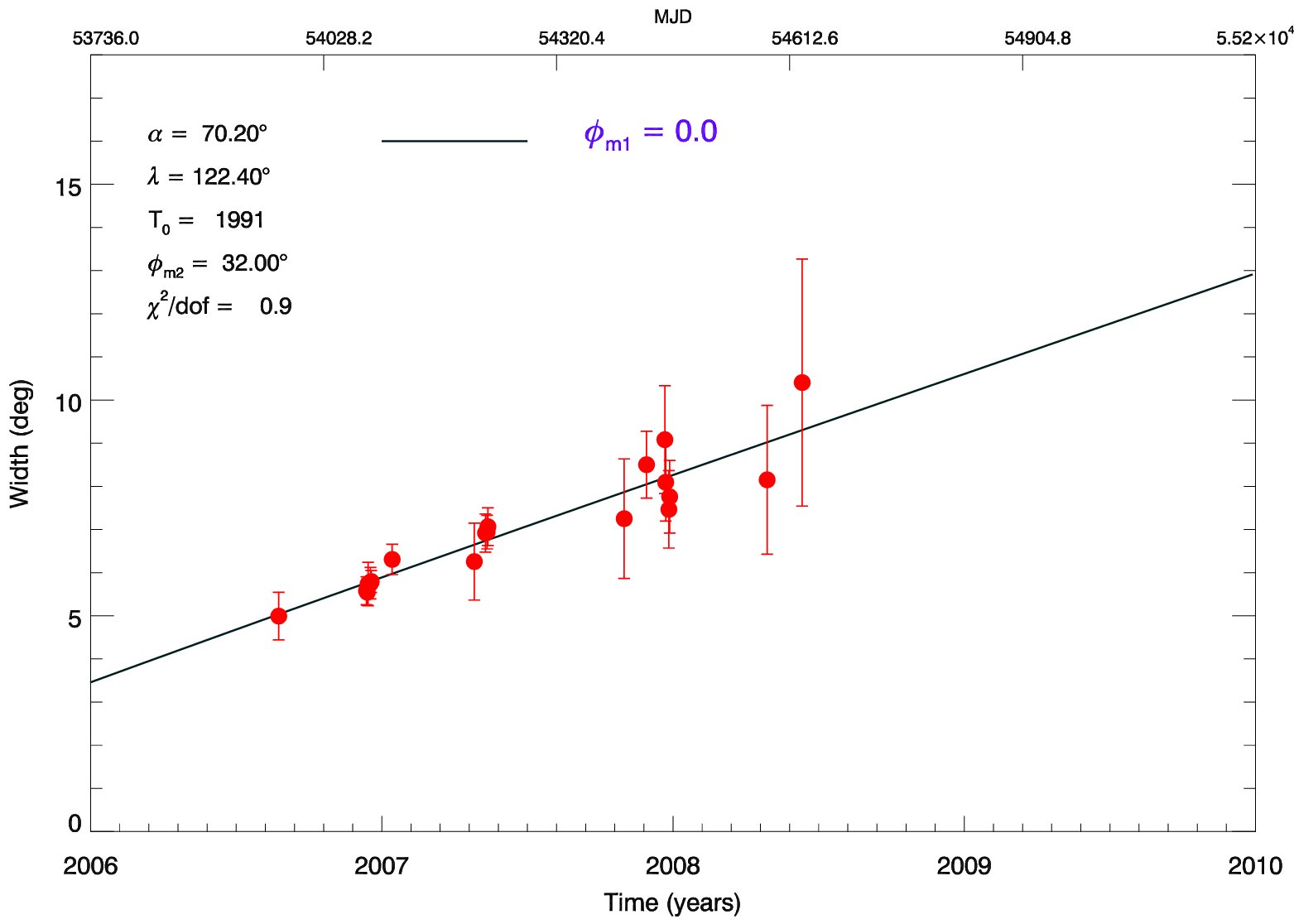}
\caption{A fit to the bp1 data taking a random point 
 within  the rhombus-shaped part of the $\chi^2$ map.}
\label{fig:fit_pair2}
\efig

Solid line in Figure \ref{fig:model-fit} presents 
 the `best' fit curve of $W(t)$ for the asymmetric outer case 
($\phi_{m1}=0$, $\phi_{m2}< 90^\circ$).  
This fit may only be nominally considered as a best solution, because it is
not statistically significant in comparison to the other models within 
the low $\chi^2$ region. The parameters of this fit 
 are: 
$\alpha = 4\degr.2$, $\lambda = 82\degr.8$, 
$\phi_{m2} = 13\degr.0$ and $T_0 = 2006$ yrs, 
 thus giving  $\chi^2 / {\rm dof} =0.9$. 
An independent fit of the symmetric beam ($\phi_{m1} = -\phi_{m2}$,
dashed line) 
 provides similar parameters, this time with $\alpha = 6\degr.6$.   
The dipole tilt seems to have suspiciously small value, however, 
by no means should these parameters be considered unique.
Satisfactory data reproduction can be achieved for any point
within the rhombus-shaped part of the $\chi^2$ map 
in Fig.~\ref{fig:contour_case3}.
An example with randomly selected $\alpha$ and $\lambda$
is shown in Fig.~\ref{fig:fit_pair2}. It has $\chi^2/{\rm dof} =0.95$. 

 Precessional
changes of $\zeta$ that correspond to the nominally-best solutions 
of Fig.~\ref{fig:model-fit}
are presented in Fig.~\ref{fig:zeta_case1}. In both cases the line of sight 
approaches the magnetic dipole axis and then retreats. This behaviour 
is not universal: other solutions (for parameters at other locations within
the rhombus) provide excellent data reproduction with the sightline
 passing far into the region on the other side of the magnetic axis.

When the fan beams are displaced to the other side of the magnetic pole
(inner traverse case, i.e.: $\phi_{m1}=180^\circ$, 
$90^\circ < \phi_{m2}< 180^\circ$), the results are analogous
to those described above. The $\chi^2$ map on the $(\alpha, \lambda)$ plane
is mirror reflected, i.e.,~the right-hand side bay in the rhombus 
of Fig.~\ref{fig:contour_case3} appears in the left-hand side corner
(i.e.~at $\alpha\la 45^\circ$, and $\lambda\sim90^\circ$). The 
 low-$\chi^2$ 
bands on the $(\phi_{m2}, T_0)$ plane get mirror-reflected and 
shift vertically by a half of precession period  
(all values of $T_0$ are therefore consistent with the data).

The pulse width data for the other bright phase (bp2)
have essentially the same character as the bp1 data.
Therefore, the same analysis applied for bp2 gave similar results, 
i.e.,~the good data match within the large part of parameter space. 
For brevity, we skip the presentation of the bp2 results.

\bfig
\centering
\fig[scale = 0.44]{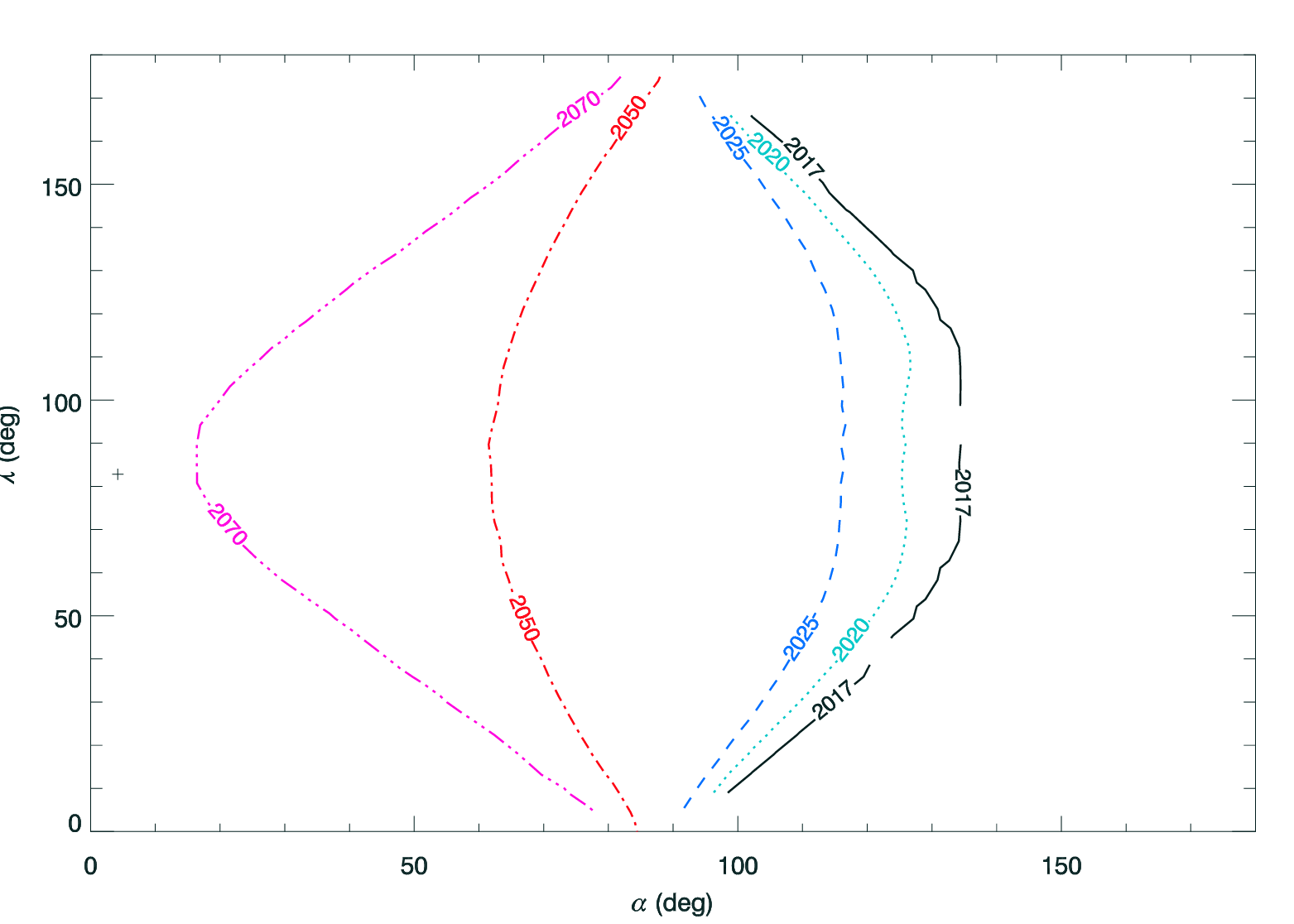}
\caption{B's reappearance time for the outer  asymmetric case of 
Fig.~\ref{fig:contour_case3}. 
 Note that the inner traverse case would produce a mirror-reflected image,
with the early reappearance time and the missing rhombus corner located 
on the left hand side. 
}
\label{reap}
\efig

\subsection{Reappearance time}

Instead of a firm prediction for the reappearance time $t_r$, 
only a map of the parameter-dependent $t_r$ can be obtained.
Fig.~\ref{reap} presents contours of fixed $t_r$ on the $(\alpha, \lambda)$
map, calculated for the case presented in Fig.~\ref{fig:contour_case3}. 
For any pair of $(\alpha, \lambda)$, values of $\phi_{m2}$ and $T_0$
are kept at their best-fit magnitudes. For each such model, 
the range of viewing angles $(\zeta_{\rm vis}^{\rm min},\zeta_{\rm vis}^{\rm max})$ 
was determined based on the 
visibility of the pulsar B in the years 2003-2008. Then $t_r$ has been 
determined as the soonest moment when $\zeta(t)$ returns back into the
visibility range.

As can be seen in Fig.~\ref{reap}, possible reappearance times 
(calculated for model parameters which can ensure good data match), 
range across most of the precession period (2017-2078). 
When moving from left to right in Fig.~\ref{reap}, 
(which was calculated for the outer traverse case), 
$t_r$ moves to earlier times. For the inner traverse 
(when beams are detectable at $\zeta < \alpha$, not shown) 
the $t_r$ contours look as a mirror reflection of Fig.~\ref{reap}, 
i.e.,~$t_r$ increases from the left-corner invisibility bay rightwards.

 It may appear disturbing that possible values of the reappearance time,
i.e. those which allow for good quality data match, span basically the 
full precession
period (Fig.~\ref{reap}), whereas the possible precession phase $T_0$ 
is excluded within a half of $P_{\rm prec}$
(Fig.~\ref{phim2t0}).\footnote{This half-period void in the allowed $T_0$ 
occurs if 
the geometry is constrained to the outer traverse case 
($|\phi_{m,1}|<90^\circ$,
$|\phi_{m,2}|<90^\circ$). Figures \ref{phim2t0} and \ref{reap} 
 were both calculated for the outer traverse geometry.} 
The variations of $W(t)$ have a sinusoid-like form, with the `wavelength' 
of the sinusoid fixed by the value of $P_{\rm prec}$. The only way 
to ensure sooner or later reappearance  
is to horizontally shift the sinusoid in $T_0$ 
(with a vertical adjustment to match the already-observed data). 
So there is
a correspondence between $T_0$ and $t_r$, but a half of the
$T_0$ values, seemingly needed to explain all $t_r$, is missing.
This apparent paradox is caused by the mathematical properties of a 
sinusoid. Let us first consider $t_r$ coincident with the last data point 
(middle 2008). That is, consider a case in which immediately after  
the maximum observed width is reached ($W=W_{\rm max}$), 
the width starts to decrease 
with no disappearance
of the B pulsar. Such behaviour would be reproduced with the maximum of the
$W(t)$ sinusoid placed at $t_r$. However, if the supposed reappearance 
is now delayed by some $\Delta t_r$
(at which time the decreasing $W$ again takes on the same maximum-observed 
value as in the middle 2008), it is enough to shift the sinusoid 
horizontally by a two times smaller time interval ($\Delta T_0= \Delta t_r/2$) 
to both pass through $W(t_r)=W_{\rm max}$ and match 
the real data points. 
A time displacement of the $W(t)$ curve produces two times larger delay 
of $t_r$. This is because the same values on both sides of the sinusoid's 
peak are twice more distant from each other than from the phase of
the peak. The nearly-half-period-long interval of $T_0$ 
in which the solution gives the correct sign of 
$dW/dt$ (see the bands in Fig.~\ref{phim2t0}) is therefore 
sufficient to ensure all possible reappearance times.

\subsection{Anticipated model constraints from future observations}

It is interesting to estimate what constraints on the model parameters will become 
available when PSR J0727$-$3039B reappears at a certain moment in 
future. To study the problem we supplemented the GBT data on $W$ 
with a set of artificial data points, created in the following way:
the original GBT data from Fig.~\ref{fig:model-fit} were reversed in time
and shifted by an arbitrary time interval. This was done because 
the precession changes the viewing angle $\zeta$ in a sinusoid-like
way, so a given range of monotonically changing $W$ is expected to 
appear twice per precession period, in a reversed time order. 
We added no data points with $W$ beyond the
 range observed so far, although the future, more
sensitive radio telescopes may be capable to detect larger range of $W$.

Such experiment shows that the constraining capability of additional data
is moderate. Only the precession phase ($T_0$) is tightly constrained down
to about a year. The dipole tilt $\alpha$ can be limited with a precision of
a few degrees, only when the B pulsar reappears close to the year 2036. In
this specific case the low-$\chi^2$ contour on the $(\alpha, \lambda)$ plane
is a narrow vertical stripe at
$\alpha \sim 90^\circ$. However for $t_r$ departing from $2036$, the
stripe bulges left or right on the 
$(\alpha, \lambda)$ plane, assuming an arc shape. 
At the same time the limits on $\alpha$ quickly become much less
precise. For $t_r\sim 2078$ or $t_r \sim2017$, the value of $\alpha$ 
(and $\lambda$) 
is basically
unconstrained. This effect is similar to the well-known
$\alpha$-$\zeta$ corellation in the polarisation angle fitting, 
which produces the 
arc-shaped $\chi^2$ contours on the $(\alpha, \zeta$) plane.
Regardless of the actual reappearance time, the additional data
do not provide any constraints on the precession angle $\lambda$, 
and the azimuthal separation of beams $\phi_{m,2}$.

It is interesting to ask if the future data allows us to discriminate
between the conal-like and the fan-beam (patchy) models.
In principle, the conal model seems to predict definite reappearance times 
(eg. $2035$ in \citealt{Perera_2010}; $2034$, $2043$, and $2066$ 
in \citealt{Lyutikov_2014}), 
so if the pulsar reappears at a different time,
only the fan beam model will stay consistent.
However, the ``conal'' model has recently gained much flexibility: 
the emission ring became a part of an ellipse with an adjustable eccentricity 
and an adjustable circumpolar azimuth of its major axis.

\section{Discussion}
\label{sec:discussion}

In terms of probability, 
the fan-shaped beam appears to be a successful model for 
the late-time evolution 
of the pulsar B profile. There is no need for fine-tuning, instead,
 there exists a large space of parameter values which 
can reproduce the observed roughly linear changes $W$. 
However, this flexibility prevents unique determination 
of pulsar and beam geometry. This ambiguity also holds for the question 
 of reappearance time of pulsar B. Since the model parameters 
are not yet fully constrained, 
the model can adjust to any date on which the pulsar may appear in future.
Such reappearance, along with new measurements of $W$ and $dW/dt$, 
would 
constrain some model parameters, provided the same radio
beam is exposed to us again.

The ambiguity in well-fitting parameters tempted us to include the 
early data on PSR J0737$-$3039B from \citet[][see Fig.~2 therein]{Burgay_2005}.
Unfortunately, for both observational and theoretical reasons, 
such analysis appears to be inconclusive.

There are the following observational problems.
First, the bp1 profiles from the period Jun 2003 -- Nov 2004 
(bottom left corner of Fig.~\ref{figdata})
cannot be 
unambiguously decomposed into two separate Gaussian components.
 Second, 
it is hard to decide if the bp1 components merge or a new (third) component
 is emerging, as suggested in \citet{Burgay_2005}. Third, the early 
bp2 profiles
seem to exhibit different shape evolution, with a nearly fixed peak-to-peak
separation, but the leading component decreasing in strength. 
This may indicate that our sightline probes widely different beam portions
in bp1 and bp2, because of orbital-dependent distortions of magnetosphere
(see \citealt{Lyutikov_2014}), or that the beam shape itself depends 
on orbital phase.

On the theoretical side, the simplest model such as the one used in the
previous section meets serious difficulties with simultaneous reproduction
of both the early (2003-Nov 2004) and late time evolution (2006-2008).
The unresolved or single shape of profile in 2005 can be interpreted as 
an overlapping emission from two nearby streams (Fig.~\ref{beamtypes}c), 
however, the large and roughly constant peak-to-peak separation 
of bp2 components in the early period cannot be understood 
if the fan beam model is limited to such two beams only. 
The conal model is even less likely to explain this, 
because of its lack of flexibility which results from the conal beam 
symmetry.
The fan beam model, on the other hand, 
still lacks the physical framework and offers some flexibility in 
choosing the beam geometry.

An important question is whether the beam responsible for the early-period 
profile 
is located on the same side of the magnetic axis 
as the late-season beam, i.e.,~whether our line of sight has passed
over the magnetic pole in 2005 (by moving meridionally, eg.~from
 the outer- to inner-traverse side)
As is the case in Fig.~\ref{beamtypes}, the system of fan beams is possibly 
different (asymmetric) on both sides of the magnetic pole (equatorward vs
poleward) which leaves lots of flexibility for the fan beam model 
to reproduce different pulse behaviour in the early and late period.
 Unfortunately, through the modelling of the late-time data from PMK10,
it is not possible to distinguish between these scenarios.
When the fan-beam model of Sect.~\ref{sec:model} is used,
 these late-time data can be well reproduced
by solutions of both types, namely those for which   
 our sightline
moves across the pole, but also those with the sightline retreating back 
after a close approach to the pole (without the pole crossing). 

\bfig
\centering
\fig[scale = 0.5]{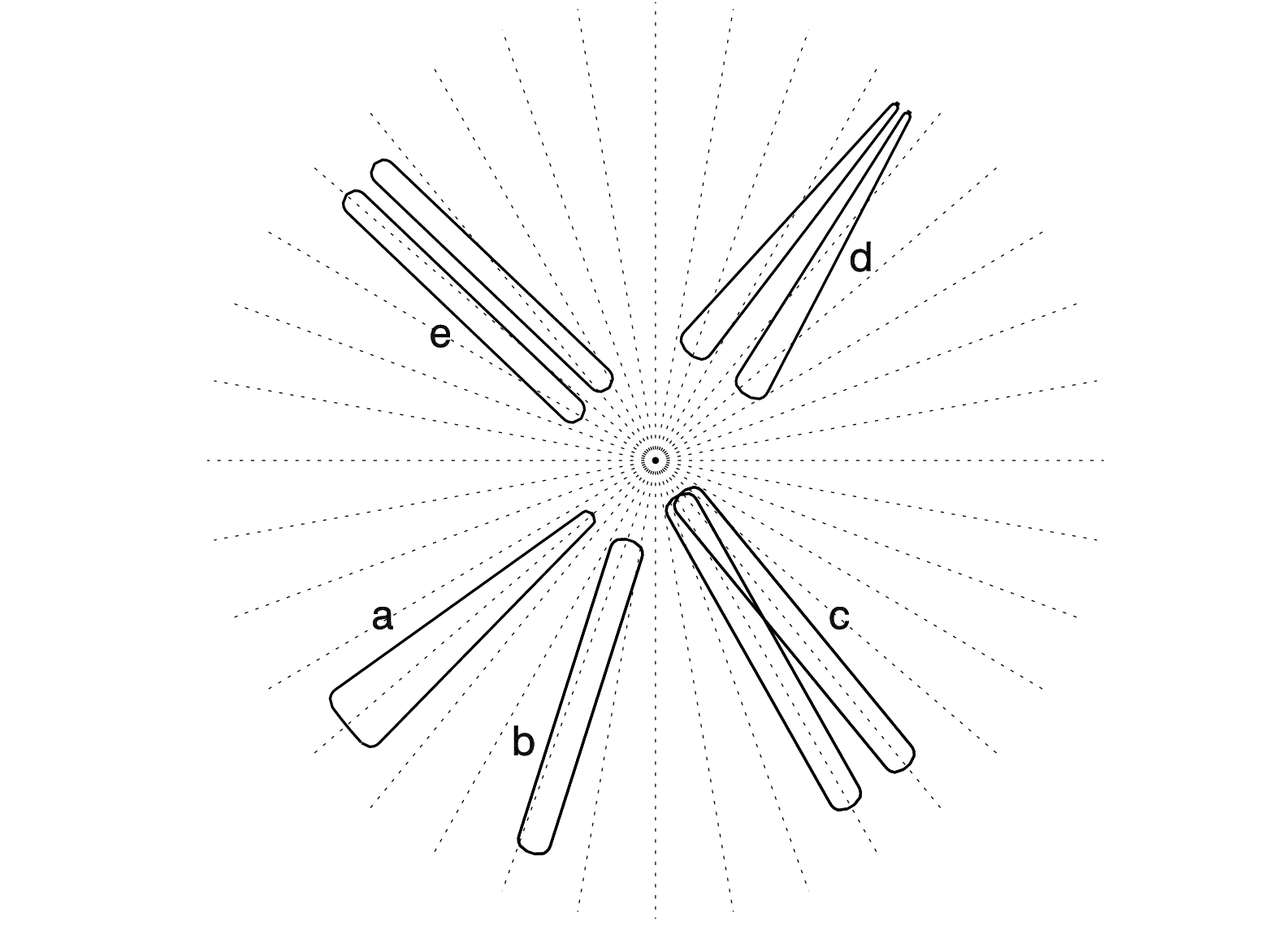}
\caption{Various types of sky-projected fan beams. 
Dotted lines mark the dipolar $B$-field, and the central dot 
is the magnetic dipole axis.
{\bf a)} Single beam which gets wider away from the axis, as caused by the
spreading $B$-field. {\bf b)} Single beam
 of a fixed width. {\bf c)} System of two fan beams which 
diverge along with the spreading $B$-field and overlap near the dipole axis.
 {\bf d)} Bifurcated beam with lobes converging away from the dipole axis.
{\bf e)} Bifurcated beam with a fixed distance between its lobes.
The latter two beams (d and e) are typical of the X-mode curvature radiation. 
Case e corresponds to a fixed radius of curvature of electron trajectory.
}
\label{beamtypes}
\efig

The modelling freedom is even larger than that, because the physical origin
of profiles' double form is not known. Aside from 
the polar-tube-related conal interpretations,
multiple fan-beam 
 types are possible, with various transverse emissivity profiles, and 
with different arrangement of fan-shaped subbeams (cartoon examples are
shown in Fig.~\ref{beamtypes}). 
Divergence of the magnetic field suggests that beams become wider at larger
distance from the dipole axis (Fig.~\ref{beamtypes}a), and this effect 
is visible in the three dimensional simulations of Wang et al.~(2014, see
figs.~5-12 therein). However, the fan beam observed in PSR J1906$+$0746 
(Desvignes et al.~2012) does not show much broadening. This is not
surprising, because the local magnetospheric emissivity depends on the 
emitted spectrum and the latter is sensitive to the energy distribution 
of emitting particles. The emitted intensity is then dependent 
on the acceleration and cooling history of the radiating charges, 
which may be different at different lateral locations in the emission
region.
In the case of the curvature radiation at standard 
conditions (charge Lorentz factor $\gamma \sim 10^2$, 
curvature radius of their trajectory $\rho\sim 10^7$ cm) the curvature
radiation spectrum barely reaches the observed $\sim 1$ GHz band. Therefore, 
any transverse differences in acceleration or cooling can influence the
observed width of the fan beam in a latitude-dependent way. 
If the emission region has the form of a dense curved stream of particles 
(such as shown in Fig.~12 of Dyks \& Rudak 2015) then 
the association of plasma density with the emitted frequency 
(also present in the radius-to-frequency mapping) 
will move
the site of strong radio emission to the center of the stream (at increasing 
altitude).
 It is for such observational and theoretical reasons, that the
latitude-dependence of the fan-beam width should be considered 
mostly unconstrained, or at least not only determined by  
the spread of the dipolar magnetic field. A fixed-width 
example of such an arbitrary beam is presented in Fig.~\ref{beamtypes}b.

Double structure of the observed profile may have various reasons.
It may be caused by emission from two, mostly independent, nearby streams, 
producing a system consisting of two fan beams (Fig.~\ref{beamtypes}c).
Alternatively, it may result from a single stream
which itself emits a split-fan beam (Fig.~\ref{beamtypes}d).
In this case the bifurcation may have either the 
micro- or macroscopic origin. 
The curvature radiation in the extraordinary mode is a known mechanism which
produces this type of a beam (see Fig.~3 in Gil et al.~2004). \nocite{glm04}
The angular separation 
of emission lobes in the beam is proportional to
$\rho^{-1/3}$ (eq.~3 in Dyks \& Rudak 2012), \nocite{Dyks_2012}
so in the dipolar field 
it slowly decreases with the distance from the dipole axis, 
as marked in Fig.~\ref{beamtypes}d. If the changes of $\rho$ 
along the particle trajectory are negligible, the beam \ref{beamtypes}e 
is expected.
In the macroscopic case 
the split-fan form may result 
from the aforementioned transverse density profile of the stream 
(Fig.~12 in \citealt{Dyks_R_2015}). 
A shallow bifurcation could possibly appear also for a stream in which most plasma is gyrating 
at a preferred pitch angle.

Given all these uncertainties (the limited time span and quality of data,
the large space of well fitting parameters, 
the model flexibility in terms of the number and location of fan
beams, and the uncertain fan beam origin and geometry), more data for a
larger interval of precession phase is required to answer the question on
whether the fan beams can explain the apparently different pulse 
behaviour observed in 2003 and 2007.

\bfig
\centering
\fig[scale = 0.5]{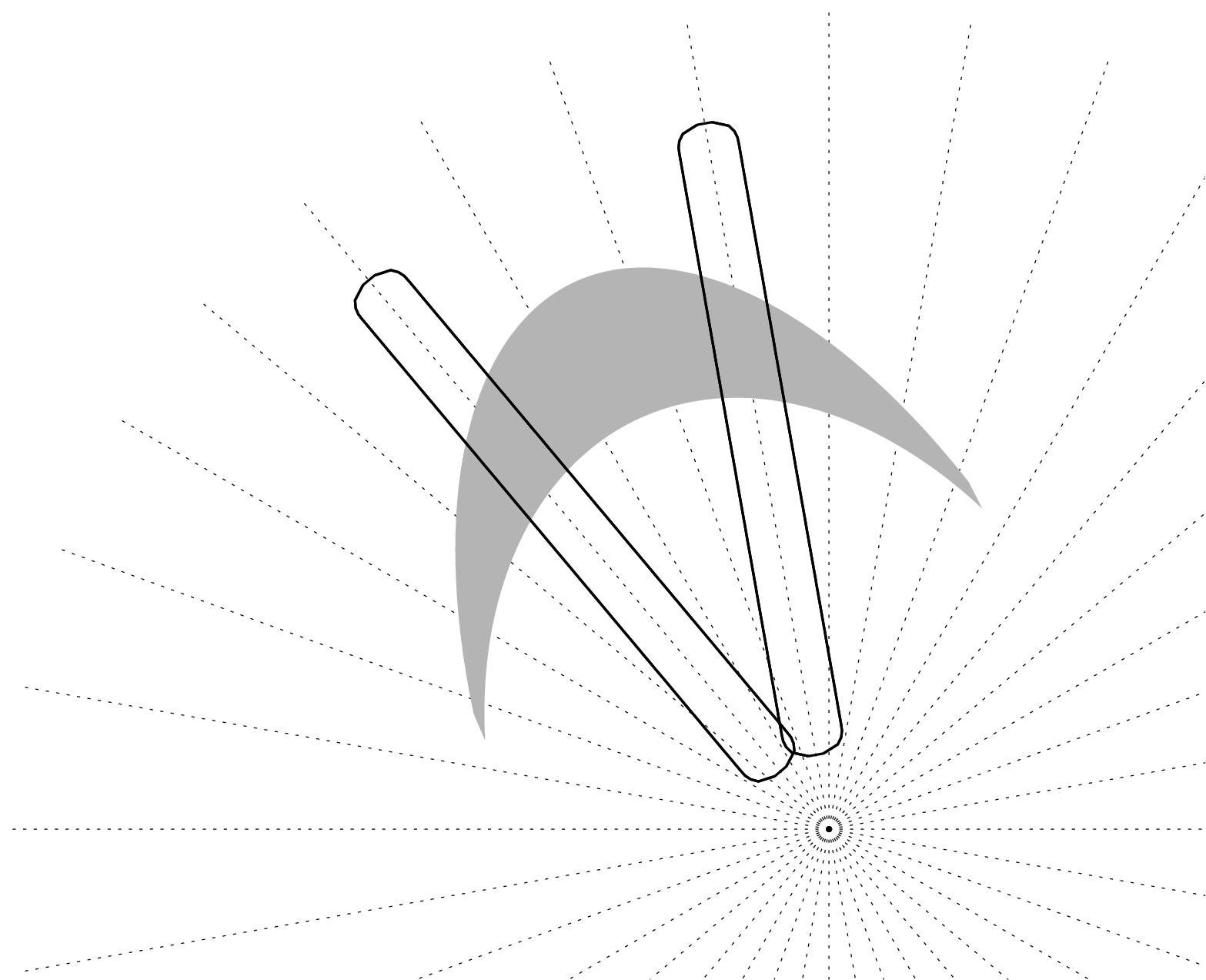}
\caption{A cartoon comparing the horseshoe beam fitted 
by Lomiashvili \& Lyutikov (2014; grey arc) to a hypothetical 
system of two fan beams (solid line contours). Because of roughly similar 
 location within the magnetosphere, the beams are suspected to 
have similar orbital visibility. Since open sides of the beams point in
opposite directions, the observed increase in $W$ requires 
an opposite precession direction for each beam type.
}
\label{lomlyutikov}
\efig

This paper has made use of the `static-shape' dipolar
magnetic field, undistorted by the pulsar A's wind. 
Such field geometry 
cannot be used to model the visibility of PSR J0737$-$3039B 
in specific orbital phase intervals. 
\citet{Lyutikov_2014}
carefully modelled
the system with a wind-distorted field. They show that 
the orbital visibility can be reproduced within a unique interval of 
precession phase, although the precession direction needs to be opposite
to that predicted by the general relativity. Their result is based on a beam
of 
 a modified-cone type: it is a piece of elliptic arc resembling 
a horseshoe, with its open side directed towards the magnetic axis. 
Let us consider the case in which the horseshoe beam  
(grey contour in Fig.~\ref{lomlyutikov}) 
is replaced with 
a V-shaped fan beam system considered 
in this paper  
(solid line contours in Fig.~\ref{lomlyutikov}). 
Let the fan beam be located within
a similar region of magnetosphere as the horseshoe beam 
(on the same side of the magnetic axis, within a
similar interval of magnetic azimuths and colatitudes), except the beam's
 open side is now pointing away from the dipole axis. 
Because of the similar magnetospheric location, the fan beam
should still be capable of reproducing the orbital visibility. 
However, since the open side of
the fan beam is pointing away from the dipole axis, the observed 
increase in $W$ can only be reproduced for an opposite precession direction
(consistent with the general relativity).   
It is therefore reasonable to expect that the V-shaped beam pointing 
towards
the dipole axis (and directing its open side away from it) can reproduce the
bright orbital phases in a more natural way than the conal-like 
horseshoe beam (i.e.,~in consistency with the general relativity).

\citet{Lyutikov_2014} have shown that the radio beam geometry 
is not very
sensitive to the wind distortion of the magnetosphere: parameters of their 
horseshoe beam are consistent with those found by PMK10  
for the static shape (pure) dipole. This suggests that our inferences on the
 performance of the fan beam geometry should prevail in the case of the
wind-distorted magnetic field.

\section{Summary}\label{sec:conclusion} 
This study shows that the 
late-time secular changes of pulsed emission from
PSR J0737-3039B 
can be explained in the framework of 
the fan-beam model. 
The  estimated space of acceptable model parameters
is quite large as compared to  that for the conal-beam model. 
However,  at least some parameters should become more tightly constrained
when the information about the reappearance of 
the pulse  and more data on $W$ is provided by future observations.

It is suggested here that the radio beam of PSR J0737$-$3039 consists 
of elongated patches with mostly azimuthal orientation. The same or 
similar geometry has appeared valid for two other precessing objects 
\citep{Manchester_2010,Desvignes_2012}. It is remarkable that 
pulsars with patchy or fan beams make up for some 50\% of precessing
objects, for which beam mapping has appeared feasible so far.

\section*{acknowledgements}
The authors thank
B.~Perera and M.~McLaughlin for the GBT data, 
as well as M.~Burgay and R.~Manchester 
 for the Parkes data on PSR J0737$-$3039B. 
We also thank A.~Frankowski and our referee for 
useful comments.
This work was supported by 
the National Science Centre grant DEC-2011/02/A/ST9/00256.

\bibliographystyle{apj}
\bibliography{binary_pulsar}

\bsp    
\label{lastpage}
\end{document}